\documentclass[final]{siamltex}

\usepackage{graphicx}

\newcommand \be{\begin{equation}}
\newcommand \ee{\end{equation}}
\newcommand \ba{\begin{eqnarray}}
\newcommand \ea{\end{eqnarray}}
\newcommand \nn{\nonumber}

\title{Convective stabilization of a Laplacian moving boundary problem
with kinetic undercooling\thanks{The exact solution (\ref{mbp25}) of the linear perturbation
problem and the slow manifold (\ref{mbp33}) were briefly presented in the Physical Review
Letter \cite{meu-05}. The paper was accepted for publication 
in SIAM J. Appl. Math. on July 16, 2007.
The work of B.M. was supported by a Ph.D. position from CWI Amsterdam.}}

\author{
        Ute Ebert\thanks{U.E. works at mainly CWI, P.O.Box 94079, 1090GB Amsterdam,
        and also at Eindhoven Univ.\ Techn., The Netherlands.} \and
        Bernard Meulenbroek\thanks{B.M. has performed this work at
        CWI, P.O.Box 94079, 1090GB Amsterdam, The Netherlands.
        He is now with Delft Univ. Techn., The Netherlands.} \and
        Lothar Sch\"afer\thanks{L.S. works at Universit\"at
        Duisburg--Essen, Lotharstr.\ 1, 47048 Duisburg, Germany.}
        }

\begin{document}

\maketitle

\begin{abstract}
We study the shape stability of disks moving in an external Laplacian field in two dimensions. The problem is
motivated by the motion of ionization fronts in streamer-type electric breakdown. It is mathematically
equivalent to the motion of a small bubble in a Hele-Shaw cell with a regularization of kinetic undercooling
type, namely a mixed Dirichlet-Neumann boundary condition for the Laplacian field on the moving boundary.
Using conformal mapping techniques, linear stability analysis of the uniformly translating disk is recast
into a single PDE which is exactly solvable for certain values of the regularization parameter. We
concentrate on the physically most interesting exactly solvable and non-trivial case. We show that the
circular solutions are linearly stable against smooth initial perturbations. In the transformation of the PDE
to its normal hyperbolic form, a semigroup of automorphisms of the unit disk plays a central role. It
mediates the convection of perturbations to the back of the circle where they decay. Exponential convergence
to the unperturbed circle occurs along a unique slow manifold as time $t\to\infty$. Smooth temporal
eigenfunctions cannot be constructed, but excluding the far back part of the circle, a discrete set of
eigenfunctions does span the function space of perturbations. We believe that the observed behaviour of a
convectively stabilized circle for a certain value of the regularization parameter is generic for other
shapes and parameter values. Our analytical results are illustrated by figures of some typical solutions.
\end{abstract}



\pagestyle{myheadings} \thispagestyle{plain} \markboth{U. Ebert, B. Meulenbroek and L. Sch\"afer}
{Convective stabilization of a Laplacian moving boundary problem $\ldots$}


\section{Introduction}

\subsection{Problem formulation in physical and mathematical context}

The mathematical model considered in this paper is motivated by the physics of electric breakdown of simple
gases like nitrogen or argon \cite{ebe-97,arr-02,arr-04,meu-04,ebe-06}. During the initial 'streamer' phase
of spark formation, a weakly ionized region extends in a strong externally applied electric field. As the
ionized cloud is electrically conducting, it screens the electric field from its interior by forming a thin
surface charge layer.  This charged layer moves by electron drift within the local electric field and creates
additional ionization, i.e., additional electron--ion--pairs, by collisions of fast electrons with neutral
molecules. We here approximate the ionized and hence conducting bulk of the streamer as equipotential. In the
non-ionized and hence electrically neutral region outside the streamer, the electric field obeys the Laplace
equation. The thin surface charge layer can be approximated as an interface which moves according to the
electric field extrapolated from the neutral region onto the interface. We therefore are concerned with a
typical moving boundary problem.

Such moving boundary problems occur in various branches of physics, chemistry or biology. The most
extensively studied examples are viscous fingering observed in two-fluid flows \cite{ben-86} or the Stefan
problem of solidification from an undercooled melt \cite{rub-71}. Other physical phenomena like the motion of
voids in current carrying metal films \cite{ho-70} lead to similar mathematical models \cite{mah-96}.

We here discuss the streamer model in two spatial dimensions, where in the simplest `unregularized' version
the basic equations coincide with those describing the motion of a small bubble in a liquid streaming through
a Hele-Shaw cell \cite{tay-59,tan-87,hon-88,ent-93}, which is a special case of two fluid flow. The
unregularized streamer model has been discussed in Ref.~\cite{meu-04,meu-05}. Restriction to two dimensions
in space allows us to use standard conformal mapping techniques \cite{ben-86,how-92} to reduce the moving
boundary problem to the analysis of the time dependence of the conformal map that maps the unit disk to the
exterior of the streamer.

It is well known that unregularized moving boundary problems of this type are mathematically ill posed
\cite{how-92}, in the sense that the moving interface generically develops cusps within finite time which
leads to a breakdown of the model. To suppress such unphysical behavior, the models are regularized by
imposing nontrivial boundary conditions on the interface. For viscous fingering typically some curvature
correction to the interfacial energy is considered. For the streamer problem a mixed Dirichlet-Neumann
boundary condition can be derived \cite{meu-05,ebe-07} by analyzing the variation of the electric potential
across the screening layer. Such a boundary condition is well known from the Stefan problem, where it is
termed `kinetic undercooling'. It rarely has been considered for Hele-Shaw type problems. There are strong
hints \cite{how-92,hoh-95,rei-99,cha-03} but no clear proof that it suppresses cusp formation. In particular,
it has been shown that an initially smooth interface stays smooth for some finite time interval.

Here we consider the linear stability of uniformly translating circles in a Laplacian potential $\varphi$
that approaches a constant slope $\varphi\propto x$ far from the circle; this means that the electric field
${\bf E}=-\nabla\varphi$ is constant far from the circle. Though this field breaks radial symmetry,
uniformly translating circles are exact solutions of the regularized problem~\cite{meu-05}. But perturbations
of these circles do not simply grow or decay locally as on a planar front or on circles in a radially symmetric
force field~\cite{hoh-95,rei-99}, but they are also convected along the boundary; this convection turns out
to be a determining part of the dynamics. Though physical streamers are elongated objects frequently
connected to an electrode, the front part
of a streamer is well approximated by a circular shape. Since it is this part that determines
the dynamics, our analysis should be relevant also for more realistic shapes like fingers where no closed
analytical solutions of the regularized uniformly translating shape are known~\cite{cha-03}.
In the sequel we will use the term `streamer' to denote
the translating circles, being aware that this is a slight abuse of the term.

\subsection{Overview over content and structure of the paper}

Regularization of the streamer model introduces some parameter $\epsilon$ that measures the effective
width of the interface relative to the typical size of the ionized region. The regularized problem allows for
a class of solutions of the form of uniformly translating circles, and linear stability analysis of these
solutions can be reduced to solving a single partial differential equation. For the special case
$\epsilon=1$, the general solution of this PDE can be found analytically, as we briefly discussed
in \cite{meu-05}. The present paper is restricted to this special case as well.

The main results of the letter \cite{meu-05} are the following:
The dynamics of infinitesimal perturbations are governed by a subgroup of the automorphisms of the unit disk.
Generically, these automorphisms convect the perturbations to the back of the moving body. Initially,
perturbations might grow, but asymptotically for time $t\to\infty$, they decay exponentially. Furthermore,
this final convergence back to the unperturbed circle follows some universal slow manifold.

The present paper contains a detailed derivation, discussion and extension of the results presented in the
letter \cite{meu-05}. Furthermore, the analyticity and completeness of temporal eigenfunctions and
the Fourier decomposition of perturbations are discussed, limit cases of the dynamics are worked out
analytically and results are demonstrated in a set of figures.

In detail, the time evolution determined by a PDE is often analyzed in terms of temporal eigenfunctions.
For the present problem in a space of functions
representing smooth initial perturbations of the moving circle, no such eigenfunctions exist. They can only
be constructed if we allow for singularities on the boundary. We here find that a subset of these functions
with time dependence $e^{-n \tau}$, $n\in {\bf N}_0$, is intimately related to the asymptotic convergence
of the perturbations. These functions show singularities only at the backside of the circle, and the front
part of any smooth perturbation can be expanded in this set of functions. The spatial domain of
convergence of this expansion increases with time and asymptotically for $t\to\infty$, it covers almost the
whole streamer. In this restricted sense these eigenfunctions form a complete set.

These results dealing with infinitesimal perturbations, of course, do not imply the asymptotic stability
of the circular shape against finite perturbations. To solve this problem, the full nonlinear theory
must be considered. Nevertheless, a first hint might be gained by considering the evolution of a finite
perturbation under the linearized dynamics. Due to the conformal mapping involved, the absence of cusps
under this evolution is not a completely trivial question. We here show that for a large range of
smooth initial conditions, the shape of the streamer stays smooth under the linearized dynamics.

All the present work deals with the exactly solvable case $\epsilon=1$ whereas the physically most interesting
case is $\epsilon\ll1$. We, however, believe that the features we could identify explicitly for $\epsilon=1$,
are generic for all $\epsilon>0$.
In particular, the subgroup of automorphisms of the unit circle leads to the basic mechanism of convective
stabilization, it is for all $\epsilon>0$ intimately related to the characteristic curves of the PDE,
and it also governs the dynamics in another exactly solvable case, namely for $\epsilon=\infty$.
Furthermore, it can be shown \cite{ef07} that the temporal eigenvalues $\lambda_n(\epsilon)$ emerging from
$\lambda_n(1)=-n$, stay negative for all $\epsilon>0$, which also indicates that the circle might be
asymptotically stable for arbitrary $\epsilon>0$.

This paper is organized as follows. In Sect.~\ref{kap2} we introduce the model, and the linear stability
analysis of translating circles is carried through in Sect.~\ref{kap3}.
These two sections are extended versions of Ref.~\cite{meu-05}. Analytical
results based on the PDE of linear stability analysis are derived in Sect.~\ref{kap4},
in particular, center of mass motion, internal motion, (non)analyticity and completeness of
eigenfunctions, intermediate growth and asymptotic decay of perturbations, Fourier representation
and motion of nonanalytical points in the complex plane of the conformal map. These dynamic features
are illustrated by explicit examples in Sect.~\ref{kap5}.
The appendix contains a discussion of the case $\epsilon=\infty$.


\section{Physical model and conformal mapping approach}\label{kap2}

\subsection{The model}

\begin{figure}
\begin{center}
\includegraphics[width=6cm]{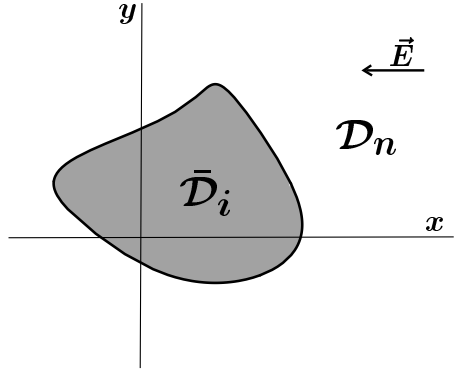}
\end{center}
\caption{Geometry of the streamer model, $\vec{E}$ is the constant far field.} \label{abb1}
\end{figure}

We assume the ionized bulk of the streamer to be a compact, simply connected domain $\bar{\cal D}_i$ of the
$(x,y)$-plane. Outside the streamer, i.e.\ in the open domain ${\cal D}_n$, there are no charges and the
electric potential obeys the Laplace equation
\begin{equation}\label{mbp1}
\Delta \varphi=0~~~\mbox{for } (x,y)\in {\cal D}_n.
\end{equation}
The streamer moves in an external electric field that becomes homogeneous far from the ionized body;
therefore the electric potential $\varphi$ at infinity obeys the boundary condition
\begin{equation}\label{mbp2}
\varphi \to E_0x+\mbox{const}~~~\mbox{for } \sqrt{x^2+y^2}\to\infty\,.
\end{equation}
This condition excludes a contribution to $\varphi$ diverging as $\ln(x^2+y^2)$ which implies that the total
charge due to the sum of all electrons and ions vanishes within $\bar{\cal D}_i$ and that the far field has the form
\begin{displaymath}
\vec{E} = -\nabla\varphi\to -E_0 \hat{\bf x}\,,
\end{displaymath}
where $\hat{\bf x}$ is the unit vector in $x$-direction. On the surface of the streamer we impose the
boundary condition
\begin{equation}\label{mbp3}
\varphi = \ell\;\hat{\bf n}\cdot\nabla\varphi\,,
\end{equation}
where $\hat{\bf n}$ is the unit vector normal to the surface pointing into ${\cal D}_n$. Here as well as in
Eq.~(\ref{mbp4}) below it is understood that the surface is approached from ${\cal D}_n$. As mentioned in the
introduction, this boundary condition results from the analysis of the variation of the potential across the
interface, and the length parameter $\ell$ can be interpreted as the effective thickness of the screening
layer. The case $\ell=0$ corresponds to the unregularized case with a pure Dirichlet condition on the moving
boundary. Dynamics is introduced via the relation
\begin{equation}\label{mbp4}
v_n=\hat{n}\cdot\nabla\varphi\,,
\end{equation}
which holds on the boundary and determines its normal velocity $v_n$. This defines our model. For further
discussion of its physical background, we refer to \cite{ebe-97,arr-02,arr-04,meu-04,ebe-06,ebe-07}.

Now obviously, $E_0$ can be absorbed into a rescaling of the potential $\varphi$ and of the time scale
inherent in the velocity $v_n$, therefore henceforth we take $E_0=1$. Clearly the model defined here is most
similar to a model of the motion of a small bubble in a Hele-Shaw cell \cite{tan-87,hon-88}, except that the
boundary condition (\ref{mbp3}) is of the form of a kinetic undercooling condition \cite{hoh-95,rei-99}.

\subsection{Conformal mapping}

A standard approach to such moving boundary problems proceeds by conformal mapping \cite{ben-86,meu-05}. We
identify the $(x,y)$-plane with the closed complex plane $z=x{+}iy$, and we define a conformal map $f(\omega
,t)$ that maps the unit disk ${\cal U}_\omega$ in the $\omega$-plane to ${\cal D}_n$ in the $z$-plane, with
$\omega=0$ being mapped on $z=\infty$
\begin{equation}\label{mbp5}
z = f(\omega ,t) = \frac{a_{-1}(t)}{\omega}+\hat{f}(\omega ,t)\,,\quad a_{-1}(t)>0.
\end{equation}
Here the function $\hat{f}$ is holomorphic for $\omega\in{\cal U}_\omega$, and we assume
that the derivatives $\partial^n_\omega$ of all orders $n$ exist on the unit circle $\partial{\cal U}_\omega$.
This restricts our analysis to smooth boundaries of the streamer. (Weaker assumptions on boundary behavior
briefly will be discussed in Sect.~\ref{ssPole}.)
We recall that the closed physical boundary can now be retrieved as $x_\alpha(t)=\Re f(e^{i\alpha},t)$
and $y_\alpha(t)=\Im f(e^{i\alpha},t))$ where the interface parametrization with the real variable
$\alpha\in[0,2\pi[$ is fixed by the conformal map.

By virtue of Eq.~(\ref{mbp1}), the potential $\varphi$ restricted to ${\cal D}_n$ is a harmonic function,
therefore it is the real part of some analytic function $\tilde\Phi(z,t)$, which under the conformal map
(\ref{mbp5}) transforms into
\begin{equation}\label{mbp6}
\Phi(\omega ,t) = \tilde{\Phi}\left(f(\omega ,t)\right) = \frac{a_{-1}(t)}{\omega}+\hat{\Phi}(\omega ,t)\,.
\end{equation}
Here the holomorphic function $\hat\Phi$ obeys the same conditions as $\hat f$ above. The pole results from
the boundary condition (\ref{mbp2}) with $E_0=1$, and (\ref{mbp5}).

Conditions (\ref{mbp3}) and (\ref{mbp4}) take
the form
\begin{eqnarray}\label{mbp7}
\left|\omega\partial_\omega f\right|\;\Re[\Phi] &=&
-\ell\; \Re\left[\omega\partial_\omega\Phi\right]\,~~~{\rm for}~\omega \in \partial{\cal U}_\omega,\\[4mm]
\label{mbp8} \Re\left[\frac{\partial_tf}{\omega\partial_\omega f}\right] &=&
\frac{\Re\left[\omega\partial_\omega\Phi\right]} {\left|\omega\partial_\omega f\right|^2}\, ~~~~~~~{\rm
for}~\omega \in \partial{\cal U}_\omega.
\end{eqnarray}
Eqs.~(\ref{mbp5}) -- (\ref{mbp8}) form the starting point of our analysis.


\section{Linear stability analysis of translating circles}\label{kap3}

\subsection{Uniformly translating circles}

A simple solution of Eqs.~(\ref{mbp7}), (\ref{mbp8}) takes the form
\begin{equation}\label{mbp9}
\left\{
\begin{array}{lll}
f^{(0)}(\omega ,t) &=&\displaystyle
\frac{R}{\omega}+\frac{2R}{R+\ell}\;t\,,\\[4mm]
\Phi^{(0)}(\omega ,t) &=&\displaystyle R\left[\frac{1}{\omega}-\frac{R-\ell}{R+\ell}\;\omega\right]\,.
\end{array}
\right.
\end{equation}
In physical coordinates $x$ and $y$, it describes circles of radius $R>0$ centered at $x(t)=v_0t$ and moving
with velocity $v_0=2R/(R+\ell)$ in direction $\hat{\bf x}$. Thus the point $\omega=1$ maps to a point at the
front and the point $\omega=-1$ maps to a point at the back of the streamer. These points will play a crucial
role in our analysis.

We note that the one-parameter family (\ref{mbp9}) of solutions parametrized by $R$, that is found in the
regularized model, is a subset of the two-parameter family found in the unregularized case $\ell=0$. As is
well known, for $\ell=0$ all ellipses with one axis parallel to $\hat{\bf x}$ are uniformly translating
solutions \cite{tay-59}.

\subsection{Derivation of the operator ${\cal L}_\epsilon$ for linear stability analysis}

We now derive the equation governing the evolution of infinitesimal perturbations of the circles
(\ref{mbp9}). In general, the parameter $R$ can become time dependent. We use the ansatz
\begin{equation}\label{mbp10}
\left\{
\begin{array}{lll}
f(\omega ,t) &=&\displaystyle
\frac{R(t)}{\omega}+x(t)+\eta\;\beta(\omega ,t)\,,\\[4mm]
{\Phi}(\omega ,t) &=&\displaystyle
R(t)\left[\frac{1}{\omega}-\frac{R(t)-\ell}{R(t)+\ell}\;\omega +\eta\;\chi(\omega ,t)\right]\,,\\[4mm]
\partial_{t}x(t)
&=&\displaystyle \frac{2R(t)}{R(t)+\ell}\,,\quad R(t)>0\,,
\end{array}
\right.
\end{equation}
where $\beta$ and $\chi$ are holomorphic functions of $\omega$ and where $\eta$ is a small parameter.
However, working to first order in $\eta$ it is found that $R$ stays
constant. This results from the fact that the dynamics embodied in Eq.~(\ref{mbp8}) strictly conserves the
area $|\bar{\cal D}_{i}|$ of the streamer, which in this context is the equivalent to the temporal
conservation of the zero order Richardson moment \cite{ent-93,how-92,ric-72}, but integrated over the
complement of ${\cal D}_n$. In terms of the mapping $f$, the conserved area $\left|\bar{\cal D}_{i}\right|$
can be written as
\begin{eqnarray}\label{mbp11}
\left|\bar{\cal D}_{i}\right| &=&
\left|\int\limits^{2\pi}_{0}d\alpha \left(\Re \left[f(e^{i\alpha},t)\right]-x(t)\right)\partial_\alpha\Im\left[f(e^{i\alpha},t)\right]\right| \nonumber \\
&=& \pi R^2(t)-\eta^{2}\int\limits^{2\pi}_{0}d\alpha\;\Re\left[\beta(e^{i\alpha},t)\right]
\partial_\alpha\Im\left[\beta(e^{i\alpha},t)\right]\,.
\end{eqnarray}
Now introducing the time independent length $R_0$ through $\left|\bar{\cal D}_{i}\right| =\pi R^2_0$, we find
$R(t)=R_0+{\cal O}(\eta^2)$, which proves that $R$ is time independent within linear perturbation theory. In
the sequel we will use $R_0$ as our length scale, introducing
\begin{equation}\label{mbp12}
\epsilon= \frac{\ell}{R_0}~~~\mbox{ and }~~~
\label{mbp13} \tau=\frac{2}{1+\epsilon}\frac{t}{R_0}\,,
\end{equation}
and rescaling $f$ and $\Phi$ by factors $1/R_0$. We note that within a dimensionless time interval
$\tau$ of order unity, the streamer moves a distance of the order of its size.

With the thus simplified ansatz~(\ref{mbp10}), Eqs.~(\ref{mbp7}) and (\ref{mbp8}) evaluated to first order in
$\eta$ take the form
\begin{equation}
\left\{
\begin{array}{l}
\displaystyle
\Re\left[\omega(\partial_\omega -\partial_\tau)\beta -\frac{1+\epsilon}{2}\omega\partial_\omega\chi \right]=0\,,\\[4mm]
\displaystyle \Re\left[
\epsilon(\omega^2+1)\omega\partial_\omega\beta-(1+\epsilon)(1+\epsilon\omega\partial_\omega)\chi\right]=0\,,
\end{array}
\right.\qquad\quad{\rm for}~ \omega \in \partial {\cal U}_\omega .
\end{equation}
Since $\beta$ and $\chi$ are holomorphic for $\omega\in{\cal U}_\omega$, these equations imply
\begin{equation}\label{mbp14}
\left\{
\begin{array}{l}
\displaystyle
\omega(\partial_\omega-\partial_\tau)\beta -\frac{1+\epsilon}{2}\omega\partial_\omega\chi=0\,,\\[4mm]
\displaystyle
\epsilon(\omega^2+1)\omega\partial_\omega\beta-(1+\epsilon)(1+\epsilon\omega\partial_\omega)\chi=ia(t)\,,
\end{array}
\right.\qquad\quad{\rm for}~ \omega \in {\cal U}_\omega,
\end{equation}
where $a(t)$ is some real function of time. $\chi$ is eliminated by substituting the expressions
for $\partial_\omega\chi$ and $\partial_\omega^2\chi$ from the first equation and its derivative
into the second equation differentiated with respect to $\omega$. This yields
\begin{equation}\label{mbp16}
{\cal L}_\epsilon\beta=0\,,
\end{equation}
where ${\cal L}_\epsilon$ is the operator
\begin{equation}\label{mbp17}
{\cal L}_\epsilon=\frac{\epsilon}{2}\;\partial_\omega\;(\omega^2-1)\omega
\;\partial_\omega+\epsilon\;\omega\partial_\omega\partial_\tau
+(1+\epsilon)\;\partial_\tau-\partial_\omega\,.
\end{equation}

\subsection{Normal form of 
${\cal L}_\epsilon$ and induced automorphisms of the unit disk}

It is instructive to transform ${\cal L}_\epsilon$ to the normal form of a hyperbolic differential operator.
We introduce \be\label{mbp18} {T}=\tanh \frac{\tau}{2}\,, \ee mapping the time interval $\tau\in[0,\infty[$
to $T\in[0,1[$, and \be \label{mbp19} \zeta = \frac{\omega +{T}}{1{+}\omega{T}}, \ee to find
\begin{eqnarray}\label{mbp20}
{\cal L}_\epsilon &=& \epsilon h(\zeta ,{T})\partial_{T}\partial_\zeta +\frac{\partial h(\zeta
,{T})}{\partial {T}}\partial_\zeta+(1+\epsilon)\partial_{T}\,,
\\[4mm]
\label{mbp21} &&h(\zeta ,{T})=\frac\omega{\partial_\zeta\omega} =\frac{(\zeta -{T})(1-{T}\zeta)}{1-{T}^2}\,.
\end{eqnarray}
This identifies the manifolds ${T}=\mbox{const}$ or $\zeta=\mbox{const}$ as the characteristic manifolds of
our problem for all $\epsilon\neq 0$.

As function of the `time-like' parameter ${T}$, $0\le {T}<1$, the transformation $\zeta=\zeta(\omega ,{T})$
in Eq.~(\ref{mbp19}) represents a semigroup of automorphisms of the unit disk, with fixed points
\begin{displaymath}
\zeta = \omega = \pm 1\,.
\end{displaymath}
For ${T}{\to}1$, corresponding to $\tau{\to}\infty$, all points $\omega\neq-1$ are mapped into $\zeta=+1$, so
that the large time behavior of any perturbation is governed by this attractive fixed point.

\subsection{Analytical solutions of Eq.~(\ref{mbp16})
for special values of $\epsilon$}

The general solution of Eq.~(\ref{mbp16}) can be found analytically for the special values $\epsilon=0$,
$\epsilon=\pm1$ and $\epsilon=\infty$. In the unregularized case $\epsilon=0$, evidently any function
\begin{displaymath}
\beta(\omega ,\tau)=\tilde{\beta}(\omega +\tau)
\end{displaymath}
is a solution, and any singularity of $\tilde{\beta}$ found in the strip
$$
0<\Re[\omega] <\infty,~~~ -1\le\Im[\omega]\le 1
$$
will lead to a breakdown of perturbation theory within finite time. This is the fingerprint of the
ill-posedness of the problem for $\epsilon=0$.

For $\epsilon= -1$, $\beta(\omega ,\tau)$ generically for all $\tau >0$ has a logarithmic singularity at
$\omega = -{T}(\tau)$. We recall that negative values of $\epsilon = \ell/R_0$ imply negative thickness of
the screening layer and thus are of no physical interest.

The case $\epsilon =+1$ is discussed in detail in the remainder of the paper. Though a regularization length
$\ell$ identical to the object size $R_0$ is somewhat artificial, it is accessible to rigorous analytical
treatment and, as explained in Section 1.2, we expect it to reveal generic features of the behavior for
all $\epsilon>0$.

This is supported by the results for $\epsilon=\infty$ which show essentially the same features as the
results for $\epsilon=1$ below. Though the limit $\epsilon\to\infty$ is physically absurd when applied to
streamers, it is worth studying with respect to the properties of the operator ${\cal L}_\epsilon$, and we
present a short discussion in the appendix.


\section{Strong screening: analytical results for $\epsilon =1$}\label{kap4}

\subsection{Analytical solution of the general initial value problem}

With the form (\ref{mbp20}) of ${\cal L}_\epsilon$, the PDE (\ref{mbp16}) for $\epsilon=1$ reduces to
\begin{equation}\label{mbp22}
\partial_{T}\;\Big(\,2+h(\zeta ,{T})\partial_\zeta\,\Big)\;\beta=0\,,
\end{equation}
showing that the function
\begin{equation}\label{mbp23}
G(\zeta)=\left(2+h(\zeta ,{T})\partial_\zeta\right)\,\beta
\end{equation}
is independent of ${T}$. To determine $\beta$, we use Eq.~(\ref{mbp21}): $h(\zeta ,{T})=\omega/\partial_\zeta
\omega$ to find
\begin{equation}\label{mbp24}
(2+\omega\partial_\omega)\;\beta(\omega ,\tau) = G(\zeta),~~~~~~ \zeta=\zeta(\omega,{T}(\tau))\,.
\end{equation}
The solution regular at $\omega=0$ takes the form
\begin{equation}\label{mbp25}
\beta(\omega ,\tau)=\int\limits^{\omega}_{0}\frac{x\;dx}{\omega^2} \;
G\left(\frac{x+{T}(\tau)}{1+x{T}(\tau)}\right)\,.
\end{equation}
A second independent solution is singular in $\omega=0$: \be \beta_{\rm
sing}(\omega,\tau)\equiv\frac1{\omega^2}. \ee The function $G$ in the regular solution (\ref{mbp25}) is
determined by the initial condition $\beta(\omega,0)$ through \be \label{G}
G(\omega)=(2+\omega\partial_\omega)\;\beta(\omega ,0). \ee It thus is holomorphic for $\omega$ in the unit
disk ${\cal U}_\omega$ and all derivatives exist on $\partial{\cal U}_\omega$, since we assume the initial
surface to be smooth. Eq.~(\ref{mbp25}) then shows that $\beta(\omega ,\tau)$ inherits these properties for
all $\tau <\infty$.

\subsection{Automorphism of unit disk and a bound on the perturbation}

It is now clear that the automorphisms $\zeta (\omega ,{T})$ of ${\cal U}_\omega$ from Eq.~(\ref{mbp21})
contain the basic dynamics and, as shown in the appendix, this also holds for $\epsilon=\infty$. This is to
be contrasted to the unregularized case $\epsilon=0$, where the dynamics amounts to a translation of the unit
disk. With the present dynamics, in the course of time larger and larger parts ${\cal U}(\delta)$ of the unit
disk ${\cal U}_\omega$ are mapped to an arbitrarily small neighbourhood $|\zeta-1|<\delta$ of the attractive
fixed point $\zeta = 1$. According to Eqs.~(\ref{mbp25}) and (\ref{G}), the initial condition in the
neighbourhood $|\omega-1|<\delta$ then determines the evolution of $\beta(\omega,\tau)$ in all ${\cal
U}(\delta)$. As a consequence, any pronounced structure found initially near $\omega_0$,
$|\omega_0-1|>\delta$, is convected towards $\omega=-1$. Quantitatively this behavior is embodied in
Eq.~(\ref{mbp35}) below, and explicit examples will be presented in section~\ref{kap5}, see, in particular,
figure~\ref{abb7}$b$.

For the further discussion we normalize $G(\omega)$ so that
\begin{equation}\label{mbp26}
\max_{|\omega|=1}|G(\omega)|=1\,.
\end{equation}
Eqs.~(\ref{mbp25}), (\ref{mbp26}) yield a bound on $\beta(\omega ,\tau)$:
\begin{equation}\label{mbp27}
\left|\beta(\omega ,\tau)\right|\le\frac{1}{2}\,;\quad |\omega |\le 1\,,\quad 0\le {T}\le 1\,.
\end{equation}
Thus the perturbation can shift the position of the streamer at most by $\eta/2$, and therefore it cannot
affect the asymptotic velocity of the propagation.

\subsection{Center of mass motion for $0\le\tau<\infty$}

In precise terms the position of the streamer can be defined as the center of mass
\begin{equation}\label{mbp28}
z_{\rm cm}=x_{\rm cm} + iy_{\rm cm} = \frac{1}{|\bar{\cal D}_i|} \int\limits_{{\cal D}_i}dx\, dy\;(x+iy)\,,
\end{equation}
where the integral is related to the first order Richardson moment. Evaluating Eqs.~(\ref{mbp28}) and
(\ref{mbp25}), we find to first order in $\eta$
\begin{eqnarray}\label{mbp29}
z_{\rm cm}&=& \tau +\eta\;\beta(0,\tau),\\
\label{cm} &&\beta(0,\tau)=\frac{G({T}(\tau))}2\,.
\end{eqnarray}
Here $\tau$ is the uniform translation of the unperturbed circle. The additional center of mass motion
(\ref{cm}) for all times is explicitly given by the initial condition $\beta(\omega,0)$ through Eq.~(\ref{G})
and the transformed time variable $T(\tau)$ from Eq.~(\ref{mbp18}); for $\tau\to\infty$, it approaches
$\beta(0,\tau)\to G(1)/2$.

\subsection{Internal motion: convergence along a universal slow manifold
for $\tau\to\infty$}

We now concentrate on the perturbation of the circular shape, given by
\begin{equation}\label{mbp30}
\tilde{\beta}(\omega ,\tau)=\beta(\omega ,\tau)-\beta(0,\tau)\,.
\end{equation}
The explicit expression
\begin{equation}\label{mbp31}
\tilde{\beta}(\omega ,\tau)=\int\limits^{1}_{0} d\rho \;\rho\left[G\left(\frac{\rho\omega +{T}}{1+\rho\omega
{T}}\right)-G({T})\right]
\end{equation}
yields
\begin{equation}\label{mbp32}
\lim_{\tau\to\infty}\tilde{\beta}(\omega ,\tau)=0
\end{equation}
for arbitrary $G$, i.e.\ for arbitrary initial condition (\ref{G}). Thus the shape perturbation converges to
zero as $\tau\to\infty$, and the circular shape is linearly stable.

We note that this holds despite the fact that the limits $\omega\to -1$ and $\tau\to\infty$, (i.e.\ ${T}\to
1$), do not commute
\begin{eqnarray*}
\lim_{{T}\to 1} \lim_{\omega\to -1} G(\zeta(\omega, {T}))=G(-1)\,, \\
\lim_{\omega\to -1} \lim_{{T}\to 1} G(\zeta(\omega, {T}))=G(+1)\,.
\end{eqnarray*}
This peculiar behavior near the backside of the streamer, at $\omega =-1$, shows up only in the rate of
convergence.

Investigating the rate of convergence for $\tau\to\infty$, we first exclude a neighborhood of $\omega =-1$
and expand $G$ in the integral (\ref{mbp31}) as \ba G\left(\frac{\rho\omega +{T}}{1+\rho\omega
{T}}\right)&=&G({T})+(1-{T}^2)\;\frac{\rho\omega}{1+\rho\omega {T}}\;G'({T})+{\cal O}(1-{T}^2)^2\,,
\nn\\
&&\mbox{where }G'(\omega)=\partial_\omega G(\omega)\,. \nn \ea With
\begin{displaymath}
1-{T}^2=4e^{-\tau}+{\cal O}(e^{-2\tau})\,,
\end{displaymath}
the integral yields
\begin{equation}\label{mbp33}
\frac{\tilde{\beta}(\omega ,\tau)}{G'(1)}=\frac{4}{\omega^2}
\left[\ln(1+\omega)-\omega+\frac{\omega^2}{2}\right]e^{-\tau}+{\cal O}(e^{-2\tau})\,,
\end{equation}
valid for
\begin{displaymath}
|1+\omega |\gg|\omega|e^{-\tau}\,.
\end{displaymath}
Thus outside the immediate neighborhood of $\omega =-1$, the shape for all smooth initial conditions with
$G'(1)\neq0$ convergences exponentially in time as $e^{-\tau}$ along a universal path in function space,
given in Eq.~(\ref{mbp33}). For $G'(1)=0$ the first non-vanishing term in the expansion of $G$ dominates the
convergence.

To analyze the neighbourhood of $\omega =-1$ we take the limit $\tau\to\infty$, with
\begin{equation}\label{mbp34}
{s}=(1+\omega)e^{\tau}
\end{equation}
fixed. We find
\begin{eqnarray}\label{mbp35}
\frac{\tilde{\beta}(\omega ,\tau)}{G'(1)} &=&
4\left(\ln(2+{s})-\tau\right)e^{-\tau} \nonumber \\
&& +\Bigg\{2G'(1)+4\ln\left(\frac{2+{s}}{4}\right) \left(G'\left(\frac{{s} -2}{{s} +2}\right)-G'(1)\right)
\nn\\
&&~~~~~+\left(2+{s}\right)\left(G(1)-G\left(\frac{{s} -2}{{s} +2}\right)\right)
-4\int\limits^{4/(2+{s})}_{0}dy\ln y\,G''(1-y)\Bigg\} \frac{e^{-\tau}}{G'(1)}
\nn\\
&&+{\cal O}\left(\tau e^{-2\tau}\right)\,.
\end{eqnarray}
In terms of $\omega$, the first contribution on the r.h.s.\ takes the form
\begin{displaymath}
4\left(\ln(2+{s})-\tau\right)e^{-\tau}=4e^{-\tau}\ln\left(2e^{-\tau}+1+\omega\right)\,,
\end{displaymath}
which shows that a logarithmic cut of $\tilde{\beta}(\omega ,\tau)$ reaches $\omega=-1$ for $\tau\to\infty$,
but with a prefactor vanishing exponentially in that limit. We thus have found a week anomaly of the
asymptotic relaxation near $\omega =-1$: In a spatial neighborhood of order $e^{-\tau}$ the exponential
relaxation is modified by a factor $\tau$. Furthermore, as mentioned above, all the initial structure of
$\beta(\omega ,0)$ is compressed into that region. This is obvious from the occurence of $G\left(\frac{{s}
-2}{{s} +2}\right)$ etc., in Eq.~(\ref{mbp35}).

To summarize, we have found that the shape of the interface for $\tau\to\infty$ converges to the circle along
a universal slow manifold (\ref{mbp33}), except for a weak anomaly (\ref{mbp35}) at the backside at
$\omega=-1$.

\subsection{(Non-)analyticity of temporal eigenfunctions}

\label{sec4.6}

In many cases, a full dynamical solution for arbitrary initial values cannot be found, and rather temporal
eigenfunctions are searched for. However, in the present problem, functions $\beta(\omega ,\tau)$ resulting
from smooth initial conditions cannot exhibit exponential behavior in time for all $\tau$, $0\le \tau
<\infty$. This is seen easily by introducing \be
G(x) = \hat G\left(\frac{x-1}{x+1}\right), \ee writing $G(\zeta)$ in the equivialent form \be
G\left(\frac{\omega +{T}}{1+\omega {T}}\right) = \hat{G}\left(\frac{\omega -1}{\omega+1}\;
e^{-\tau}\right)\,, \ee and substituting this form into Eq.~(\ref{mbp25}). Postulating strict exponential
time behavior $\beta\sim e^{-\lambda\tau}$ one finds
\begin{equation}\label{mbp36}
\beta(\omega ,\tau) \propto e^{\lambda\tau}\;\beta_\lambda(\omega),~~~~~~ \beta_\lambda(\omega)=
\int\limits^{1}_{0}d\rho\rho\left(\frac{\omega\rho -1}{\omega\rho +1}\right)^{\lambda }\,.
\end{equation}
Any eigenfunction $\beta_\lambda(\omega ,0)$ with $\lambda\ne0$ clearly is singular either at $\omega =+1$,
or at $\omega =-1$, or at both points. It therefore conflicts with smooth initial conditions. On the other
hand, omitting a neighbourhood of $\omega=-1$, eigenfunctions exist for all $-\lambda\in{\bf N}_0$.

\subsection{Completeness of the eigenfunctions near $\omega=1$}

In some neighborhood of $\omega=1$, we even can show that any regular solution $\beta(\omega ,\tau)$ can be
expanded in terms of the `eigenfunctions' $\beta_{-n}(\omega)$, $n\in {\bf N}_0$. This results from the
Taylor expansion
\begin{equation}\label{mbp36b}
\hat{G}(y)= \sum^{\infty}_{n=0} \hat{g}_n y^n,
\end{equation}
which by assumption converges in a disk of radius $\hat{r}>0$: Rewriting Eq.~(\ref{mbp25}) as \ba
\beta(\omega,\tau) &=&\int_0^1 \frac{x\;dx}{\omega^2}\;G\left(\frac{x+T}{1+x T}\right)- \int_\omega^1
\frac{x\;dx}{\omega^2}\;G\left(\frac{x+T}{1+x T}\right)
\nn\\
&=& \frac{M(T)}{\omega^2}-\sum_{n=0}^\infty \hat g_n\; e^{-n \tau} \int_\omega^1 \frac{x\;dx}{\omega^2}\;
\left(\frac{1-x}{1+x}\right)^n \ea and $\beta_{-n}(\omega)$ in a similar form as \be
\beta_{-n}(\omega)=\frac{M_n}{\omega^2}- \int_\omega^1 \frac{x\;dx}{\omega^2}\;
\left(\frac{1-x}{1+x}\right)^n, \ee we find \be \beta(\omega,\tau)=\frac{M(T)}{\omega^2}+\sum_{n=0}^\infty
\hat g_n \;\left[\beta_{-n}(\omega)-\frac{M_n}{\omega^2} \right]\;e^{-n \tau}. \ee Provided $e^{-\tau}<\hat
r$, we can separate the sum into the contribution $\propto 1/\omega^2$ and the rest. Since both
$\beta(\omega,\tau)$ and the $\beta_{-n}(\omega)$ are regular at $\omega=0$, the contributions
$\propto1/\omega^2$ have to cancel, which yields the final result
\begin{equation}\label{mbp16c}
\beta(\omega ,\tau)=\sum^{\infty}_{n=0} \hat{g}_n \;\beta_{-n}(\omega) \; e^{-n\tau}\,.
\end{equation}
This result is valid for $e^{-\tau}<\hat{r}$ in the disk
\begin{displaymath}
\left|\frac{1-\omega}{1+\omega}\right| e^{-\tau} < \hat{r}\,.
\end{displaymath}
It generalizes the asymptotic result (\ref{mbp33}).
Indeed, the universal shape relaxation found in (\ref{mbp33}) together with the center of mass relaxation
(\ref{cm}) precisely follow the slowest eigenfunction from (\ref{mbp36b}) with $\lambda=-n=-1$.
Furthermore this result shows that the range of validity of the expansion (\ref{mbp16c}) increases with
$\tau$ and asymptotically covers the whole complex plane except for the special point $\omega=-1$.

\subsection{Intermediate temporal growth and coupling of Fourier modes}

Having found that the space of regular functions does not allow for strictly exponential time behavior, we
now consider the typical time variation of smooth perturbations. Before the exponential relaxation sets in,
such perturbations typically will grow, and this growth can be quite dramatic. As an illustration we consider
a perturbation defined by
\begin{displaymath}
G(\omega)=\omega^{k}\,,\quad k\gg 1\,,
\end{displaymath}
corresponding to initial conditions
\begin{equation}\label{mbp37}
\beta(\omega ,0)=\frac{\omega^k}{k+2}\,.
\end{equation}
For ${T}=1-e^{-\theta}/k$, corresponding to times $\tau = \theta+\ln (2k)+{\cal O}(1/k)$, we can write
\begin{displaymath}
G\left(\frac{\omega +{T}}{1+\omega {T}}\right) =
\left(\frac{1-\frac{e^{-\theta}}{1+\omega}\frac{1}{k}}{1-\frac{\omega
e^{-\theta}}{1+\omega}\frac{1}{k}}\right)^k = \exp\left[-e^{-\theta}\frac{1-\omega}{1+\omega}\right]
\left(1+{\cal O}\left(\frac{1}{k}\right)\right)\,,
\end{displaymath}
where we again excluded some neighborhood of $\omega =-1$. Substituting this expression into
Eq.~(\ref{mbp31}) we find on the unit circle $\omega=e^{i\alpha}$: \ba\label{mbp38}
&&\tilde{\beta}(e^{i\alpha},\tau)
\\
&&= \int\limits^{1}_{0}d\rho\rho
\exp\left[-e^{-\theta}\frac{1-\rho^2-2i\rho\sin\alpha}{1+\rho^2+2\rho\cos\alpha} \right] -
\frac{1}{2}\exp\left[-e^{-\theta}\right]+{\cal O}\left(\frac{1}{k}\right)\,. \nn \ea

\begin{figure}
\begin{center}
\includegraphics[width=6cm]{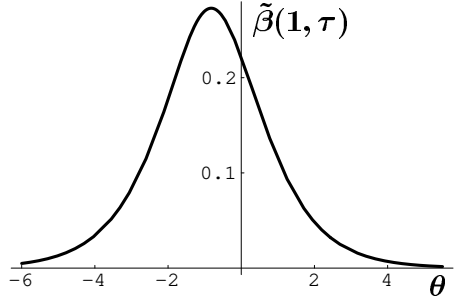}
\end{center}
\caption{$\tilde\beta(e^{i\alpha},\tau)$ from Eq.~(\ref{mbp38}) for $\alpha=0$ as a function of subtracted
time $\theta=\tau-\ln 2k$.} \label{abb2}
\end{figure}

Figure~\ref{abb2} shows this function, evaluated at $\alpha =0$ ($\omega=1$). The behavior is quite peculiar.
Up to times of order $\ln k$ the perturbation stays of order $1/k \ll 1$, then it increases roughly
exponentially up to values of order $1$, and finally it decreases again exponentially, approaching the slow
manifold (\ref{mbp33}). Thus for very large $k$ the initial perturbation $\beta(\omega ,0)\sim1/k$ in some
time interval can be amplified by a factor of order $k$, and Eq.~(\ref{mbp38}) shows that the leading
behavior in that time interval is independent of $k$.

Closer analysis shows that in terms of a formal Fourier expansion
\begin{equation}\label{mbp39}
\tilde{\beta}(e^{i\alpha},\tau)=\sum^{\infty}_{n=1}a_n(\tau) e^{in\alpha},
\end{equation}
the amplification is carried by the low modes, $n={\cal O}(1)$. As will be illustrated by an explicit example
below, cf.\ figure~\ref{abb5}$b$, in such a mode representation the time evolution feeds the strength of the
perturbation sucessively into lower and lower modes. This is equivalent to the observation that the
automorphism $e^{i\alpha}\to\zeta(e^{i\alpha},{T})$ drives all the perturbative structure towards $\alpha
=\pi$ and smoothens the remainder of the interface. Note, however, that starting with a perturbation
$\sim\omega^k$ in the course of time also modes $n>k$ are (weakly) populated to build up a complicated
structure near $\omega =-1$. We recall that for the unregularized model $\epsilon=0$, the time evolution of a
perturbation $\propto \omega^k$ populates only modes $k\le n$ \cite{meu-04}.

\subsection{Motion of the zeros of $\partial_\omega f$ and cusps}

\label{ssPole}

So far we have shown that the propagating circle is linearily stable, i.e., we implicitly considered
perturbations of infinitesimal strength $\eta$. The full nonlinear evolution of a finite perturbation is
beyond the scope of this paper. Still, it clearly is a question of practical interest, whether a finite
perturbation evolving under the linearized dynamics, for all times satisfies the assumptions underlying the
conformal mapping approach. For the mapping to stay conformal, all the zeros of $\partial_\omega f(\omega
,\tau)$ must stay outside the unit circle. We thus here analyze the roots of the equation
\begin{equation}\label{mbp40}
0 =
\partial_\omega f(\omega ,\tau)
= -\frac{1}{\omega^2}+\eta\;\partial_\omega\beta(\omega, \tau)\,.
\end{equation}
Using Eqs.~(\ref{mbp24}), (\ref{mbp25}), we can rewrite this equation as
\begin{equation}\label{mbp41}
2\eta \int\limits^1_0 d\rho\, \rho\left[G\left(\frac{\omega +{T}}{1+\omega{T}}\right)
-G\left(\frac{\rho\omega+{T}}{1+\rho\omega{T}}\right)\right] = \frac{1}{\omega}\,.
\end{equation}
With our normalization (\ref{mbp26}) of $G$, for all $\omega$ in the closed unit disk the l.h.s.\ of this
equation is bounded by $2|\eta|$. We conclude that the bound
\begin{equation}\label{mbp42}
|\eta |<\frac{1}{2}
\end{equation}
guarantees that within the framework of first order perturbation theory the mapping stays conformal for all
times. We now will show that this bound in general cannot be improved.

For $\tau\to\infty$, zeros of $\partial_\omega f(\omega ,\tau)$ reach $\omega =-1$, which is a consequence of
the fact that in this limit an infinitesimally small neighborhood of $\omega =-1$ under the mapping
$\omega\to\zeta$ is mapped essentially on the whole complex plane. We now analyze this limit for the simple
example $G(\omega)=\omega$. Substituting this form into the asymptotic behavior (\ref{mbp35}) and using the
definition (\ref{mbp34}) of ${s}$, we find
\begin{displaymath}
\partial_\omega\beta
= \frac{4}{2+{s}}+{\cal O}(\tau e^{-\tau})\,.
\end{displaymath}
Eq.~(\ref{mbp40}) reduces to ${s} = 4\eta -2$, showing that a zero $\omega_0$ of $\partial_\omega f(\omega
,\tau)$ approaches $\omega =-1$ as
\begin{displaymath}
\omega_0 = -1+(4\eta -2)e^{-\tau}\,.
\end{displaymath}
For $\omega_0$ to come from outside the unit circle we clearly must have
\begin{equation}\label{mbp43}
\Re[\eta]<\frac{1}{2}\,.
\end{equation}
To get some feeling for the estimate (\ref{mbp42}), we note that for  $G(\omega)=\omega^k$ the map initially (for
$\tau=0$) is conformal provided $|\eta|<1+2/k$. We conclude that under the linearized dynamics a large part
of smooth initial conditions relaxes to the circle.

In all the discussion of this section we have assumed the initial boundary to be smooth, so that all
derivatives $\partial^n_\omega G(\omega)$ exist on the boundary $|\omega|=1$. Inspecting the results it is
obvious that this assumption can be considerably relaxed, since only those derivatives which show up
explicitly, have to exist. Thus, for exponential relaxation (\ref{mbp33}) outside the neigborhood of
$\omega=-1$ to prevail, the existence of $\partial_\omega G(\omega)$ is sufficient, which amounts to the
condition that the curvature of the initial boundary is well defined. For the circle to be linearly stable,
Eq.~(\ref{mbp32}), it is sufficient that $G(e^{i\alpha})$ is bounded and continuous, which implies that the
boundary has a well defined slope.

If the initial boundary shows a cusp, the time evolution sensitively depends on the details. If the cusp is
found in forward direction, so that $G(\omega)$ diverges for $\omega\to 1$, the streamer will be strongly
accelerated. In a related model \cite{hon-88}, such an effect has been pointed out before. Furthermore the
shape will not relax to a circle, and the conformal map will  presumably break down at finite time. If the
cusp does not affect the analyticity of $G(\omega)$ near $\omega =1$, it is convected towards the back and
broadened, whereas the front of the streamer approaches the circular shape. Still, however, conformality of
the map may break down at finite time.


\section{Explicit examples for $\epsilon =1$}\label{kap5}

We here illustrate the general results by some examples.

\subsection{The evolution of Fourier perturbations}

\begin{figure}
\begin{center}
\includegraphics[width=10cm]{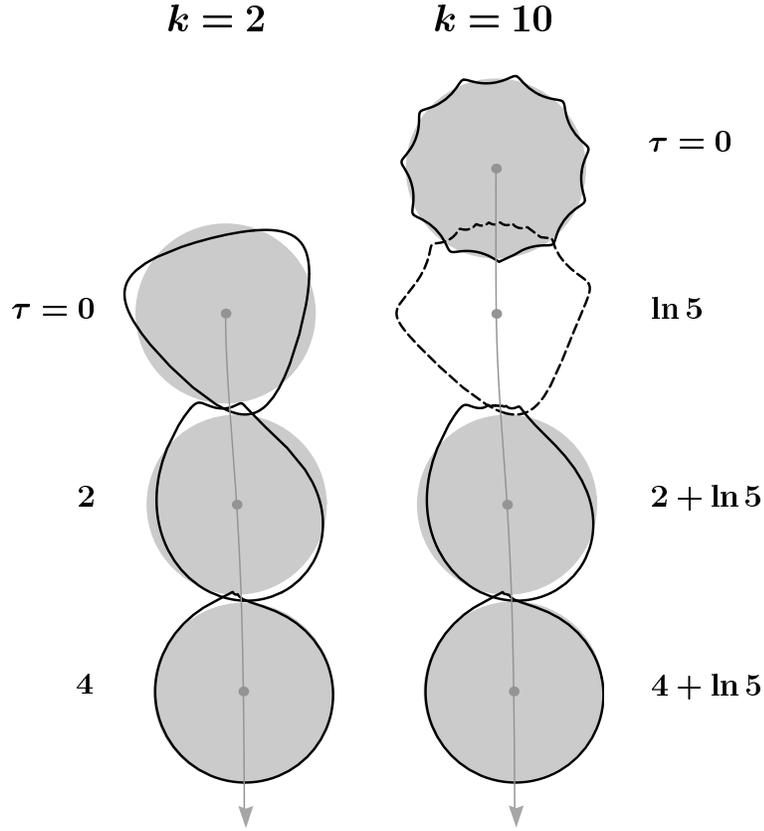}
\end{center}
\caption{Snapshots of the evolution of the streamer for $k=2$ (left column) and $k= 10$ (right column) at the
indicated instants of time. The solid lines represent the perturbed interfaces. The gray disks move with the
center of mass velocity (\ref{mbp54}) of the perturbed circles. One gray disk has been omitted for clarity.
See the text for further discussion.} \label{abb3}
\end{figure}

 We first consider perturbations of the form
\be \beta^{[k]}(\omega ,0)=\frac{\omega^k}{k+2},~~~\mbox{ i.e. } G(\omega)=\omega^k. \ee The integral
(\ref{mbp25}) is easily evaluated to yield
\begin{eqnarray}\label{mbp44}
\beta^{[k]}(\omega ,\tau) &\displaystyle=\frac{1}{2\omega^2{T}^2}\Bigg\{
&{T}^k{+}\left(({T}\omega )^2{-}1\right)\zeta^k \nonumber \\
&& +k \left(1{-}{T}^2\right) \Bigg[{T}^k{-}(\omega{T}{+}1)\zeta^k+\frac{1{+}k{+}{T}^2(1{-}k)}{T^k} \nonumber \\
&& \qquad\qquad\qquad\qquad\qquad\cdot \left(\ln (1{+}\omega{T}) -\sum^{k{-}1}_{\nu=1}
\frac{{T}^\nu}{\nu}(\zeta^\nu{-}{T}^\nu)\right)\Bigg]\Bigg\}\,,
\end{eqnarray}
where ${T}={T}(\tau)$ and $\zeta=\zeta(\omega ,{T}(\tau))$ are given by Eqs.~(\ref{mbp18}), or (\ref{mbp19}),
respectively. In figure~\ref{abb3} we have plotted snapshots of the resulting motion of the interface,
determined as
\begin{equation}\label{mbp45}
z=x+iy=\frac{1}{\omega}+\tau+\eta\;\beta^{[k]}(\omega ,\tau)\,,\quad \omega =
e^{i\alpha}\,,\quad0\le\alpha\le 2\pi\,.
\end{equation}
The direction of motion, i.e.\ the positive $x$-direction, is downwards. Together with the moving interface
we show the unperturbed circular streamer at different times as gray disks with the center moving according
to \be \label{mbp54} z_{\rm cm}(\tau)=\tau +\frac{\eta}{2}\;G({T}(\tau)) = \tau +\frac{\eta}{2}\;\tanh^k
\frac\tau2\,, \ee as predicted for the center of mass motion for the perturbed streamer in Eq.~(\ref{mbp29}).

\begin{figure}
\begin{center}
$a)$ \includegraphics[width=6cm]{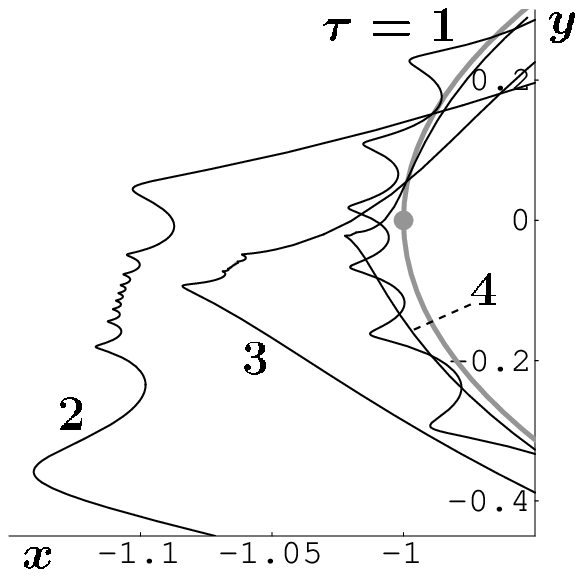} $b)$ \includegraphics[width=6cm]{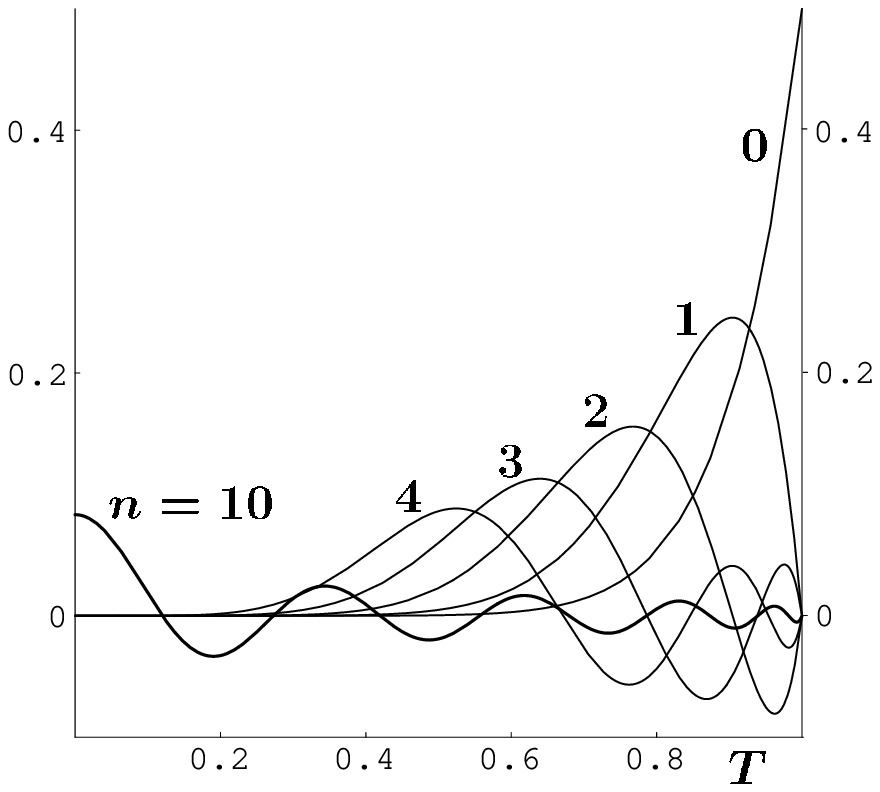}
\end{center}
\caption{$a)$ Magnified plot of the backside of the streamer for $k=10$, $\eta=0.6\;e^{i\pi/4}$ (as in the
right column of figure~\ref{abb3}) for the $\tau$ values given. The overall motion is subtracted. We observe
the compression of the fine structure and the intermediate growth of the perturbation. Asymptotically for
$\tau\to\infty$, the structure converges to the gray circle. In the comoving frame, the gray dot marks
$x+iy=-1$ which is the point to which the structure finally is contracted. Note that the scale of $x$ is
stretched compared to that of $y$, and that the figure is turned relative to figure~\ref{abb3}.
\newline
$b)$ The amplitudes $a_n$ as in Eq.~(\ref{mbp39}) as a function of $T$ for $k=10$; the values of $n$ are
given. } \label{abb4}\label{abb5}
\end{figure}

In figure~\ref{abb3} we perturbed the circle by $\eta\;\beta^{[k]}$, $k=2$ or $k=10$, using the same
parameter $\eta=0.6 e^{i\pi/4}$ in both cases. The starting position for $k=10$ is shifted relative to that
for $k=2$ by a distance corresponding to $\Delta\tau=\ln 5$. As discussed below Eq.~(\ref{mbp38}), for $1\ll
k_1< k_2$ we expect
$$
\beta^{[k_1]}(\omega ,\tau) \approx  \beta^{[k_2]}(\omega ,\tau +\ln(k_2/k_1)).
$$
Figure~\ref{abb3} illustrates that such a `universality' for the gross structure holds down to very small
$k$. (Of course the choice of differing values of $\eta$ would distort the figures and mask  this feature.)
Basically during time evolution the initial maximum closest to the forward direction is smeared out and
builds up the asymptotic circle, whereas all other structures are compressed at the backside. For $k=10$ the
complicated structure at the back is magnified in figure~\ref{abb4}$a$. Figure~\ref{abb5}$b$ shows the time
dependence of the coefficients $a_n$ of the low modes $e^{in\alpha}$ in the expansion (\ref{mbp39}), again
for $k=10$. It illustrates how the strength of the perturbation cascades downwards in $n$ and increases in
time, until it completely is absorbed into the lowest mode, i.e., the overall shift of the circle. We should
recall, however, that also modes $n>k$ are weakly populated to build up the structure at the back.

\begin{figure}
\begin{center}
\includegraphics[width=6cm]{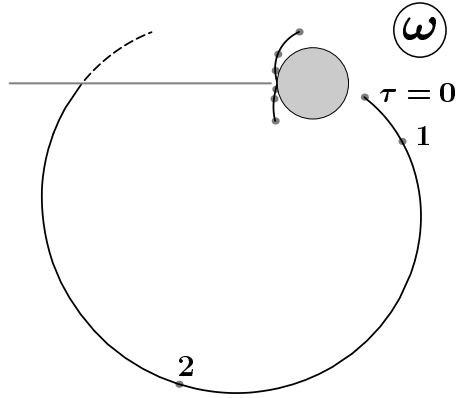}
\end{center}
\caption{Motion of the zeros of $\partial_\omega f$ in the $\omega$-plane for $k=2$ and
$\eta=0.6\;e^{i\pi/4}$ (as in the left column of figure~\ref{abb3}). The dots give the position for
$\tau=0,~1,~2$. The horizontal line is the cut for $\tau=2.51$, where one zero enters the second sheet
(broken curve). The unit disk is also shown.} \label{abb6}
\end{figure}

For $k=2$, figure~\ref{abb6} shows the motion of the zeros of $\partial_\omega f(\omega ,\tau)$ in the
complex $\omega$-plane, as discussed in section \ref{ssPole}. It corresponds to the $k=2$ part of
figure~\ref{abb3}. Two zeros, which initially are close to the backside of the unit circle, for
$\tau\to\infty$ approach $\omega=-1$. They clearly are associated with the two maxima that in the comoving
frame are convected towards $z=x+iy=-1$. The third zero, originally found close to $\omega=+1$, after a large
excursion leaves the physical sheet at time $\tau\simeq 2.51$. The logarithmic cut is on the negative axes,
with the branchpoint $\omega_{\rm bp}=-1/{T}(\tau)$ reaching $\omega=-1$ for $\tau\to\infty$.

\subsection{The evolution of localized perturbations}

\begin{figure}
\begin{center}
\includegraphics[width=10cm]{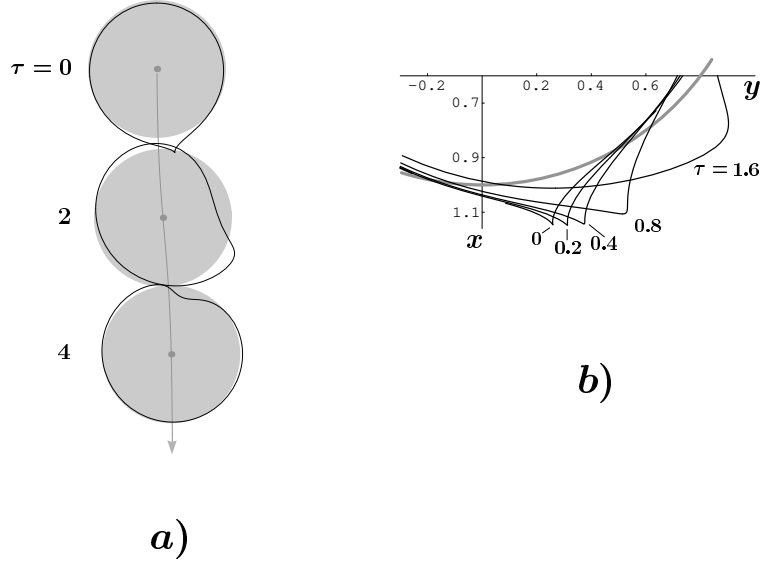}
\end{center}
\caption{$a)$ Time evolution of a localized perturbation as described in the text. $~~~$ $b)$ Evolution of
the initial peak for shorter times as indicated. The overall motion of the streamer is subtracted. A part of
the asymptotic circle is shown in gray.} \label{abb7}
\end{figure}

We finally consider some more localized perturbation, defined by \be
G(\omega)=\frac{(1-\gamma)e^{i\alpha_0}}{\omega-\gamma e^{i\alpha_0}}\,,\quad \gamma > 1\,, \ee corresponding
to an initial perturbation \be \eta\;\beta(\omega ,0)= \eta\; \frac{(1-\gamma)\gamma}{\omega^2}\;
e^{2i\alpha_0}\left[\ln\left(1-\frac{\omega}{\gamma}\; e^{-i\alpha_0}\right)-\frac{\omega}{\gamma}
\;e^{-i\alpha_0}\right]\,. \ee The result for $\beta(\omega ,\tau)$ reads \be \beta(\omega
,\tau)=\frac{(\gamma-1)e^{i\alpha_0}}{\gamma e^{-i\alpha_0}-{T}(\tau)}
\left\{\frac{{T}(\tau)}{2b(\tau)}-\left(1-\frac{{T}(\tau)}{b(\tau)}\right) \frac{\ln(1+b(\tau)\omega
)-b(\tau)\omega}{(b(\tau) \omega)^2}\right\}\,, \ee where \be b(\tau)=\frac{1-{T}(\tau)\gamma
e^{i\alpha_0}}{{T}(\tau)-\gamma e^{i\alpha_0}}\,. \ee We note that $b(\tau)\to1$ for ${T}(\tau)\to1$, so that
in the large time limit the logarithmic cut reaches $\omega=-1$. As discussed in the context of
Eq.~(\ref{mbp35}), this is a generic feature of the present problem. Our choice of parameters $\gamma=1.1$,
$\alpha_0=-\pi/12$, $\eta=1.5$, almost produces a cusp in the initial condition: the only zero of
$\partial_\omega f(\omega ,0)$ is found at $\omega_0=1.001 \exp(-.243 i)$. This zero, however, is driven away
from the unit circle and leaves the physical sheet. Another zero that entered the physical sheet somewhat
earlier, for $\tau\to\infty$ reaches $\omega=-1$. Figure~\ref{abb7}$a$ shows snapshots of the time evolution
of the perturbed interface in a representation like figure~\ref{abb3}. It illustrates how the peak rapidly is
smeared out and the interface becomes smooth. Figure~\ref{abb7}$b$ follows the evolution of the peak for
short times and shows how it is convected and broadened.

We finally note that in the special case, where the initial peak strictly points in forward direction
$(\alpha_0=0)$, convection cannot take place. The peak simply is broadened and vanishes, whereas some new
peak shows up at the back for intermediate times.


We acknowledge helpful and motivating discussions with F. Brau, A. Doelman, J. Hulshof, H. Levine, L.P.
Kadanoff, S. Tanveer and S. Thomae.


\begin{appendix}

\section{The limit $\epsilon\to\infty$}

For $\epsilon\to\infty$, the PDE (\ref{mbp16}) with the form (\ref{mbp20}) of ${\cal L}_\epsilon$ reduces to
\be \label{a1} \Big(h(\zeta,T)\;\partial_\zeta+1\Big)\;\partial_T\;\hat\beta(\zeta,T)=0, ~~~~~~\mbox{ where }
\label{a2} \hat\beta(\zeta,T)\equiv \beta(\omega,\tau). \ee Eq.~(\ref{a1}) allows for a large set of
solutions obeying the same initial condition \be \label{a3} \beta(\omega,0)=\beta_0(\omega), \ee but imposing
regularity on the unit disk ${\cal U}_\omega$, we single out the simple form \be \label{a4}
\beta(\omega,\tau)=\beta_0(\zeta). \ee Thus for $\epsilon=\infty$, the dynamics is simply given by the
automorphisms $\omega \longrightarrow \zeta(\omega,T)$. This implies that $\beta(\omega,\tau)$ is bounded
uniformly in $\tau$ as \be \label{a5} \big| \:\beta(\omega,\tau)\:\big| \le \max_{\omega\in\partial{\cal
U}_\omega} \big| \:\beta_0(\omega)\:\big|, \ee so that in contrast to the case $\epsilon=1$, there is no
intermediate growth of the perturbations.

The shift of the center of mass is given by (cf.~Eq.~(\ref{mbp29})): \be \label{a6}
\beta(0,\tau)=\beta_0(T(\tau))
= \beta_0(1)-2\;\beta_0'(1)\;e^{-\tau}+{\cal O}\left(e^{-2\tau}\right), \ee and except for the point
$\omega=-1$, the shape again converges exponentially in time to the circle along the universal slow manifold
\be \label{a7} \beta(\omega,\tau)-\beta(0,\tau)=\beta_0'(1)\;\frac{4\;\omega}{1+\omega}\; e^{-\tau}+{\cal
O}\left(e^{-2\tau}\right), \ee cf.~Eq.~(\ref{mbp33}) for $\epsilon=1$. Again the neighbourhood of $\omega=1$
for time $\tau=0$, more precisely $\beta_0(1)$ and $\beta_0'(1)$, determine the long time convergence. Since
by assumption $\beta_0(\omega)$ is analytical at $\omega=1$, evidently an eigenfunction expansion in the
sense of subsection~\ref{sec4.6} exists.

The only major difference to the case $\epsilon=1$ concerns the point $\omega=-1$. Clearly, \be \label{a8}
\beta(-1,\tau)\equiv\beta_0(-1) \ee independently of $\tau$, and indeed for $\tau\to\infty$ the conformality
of the mapping breaks down in the neighbourhood of $\omega=-1$ since
$\partial_\omega\beta(\omega,\tau)\big|_{\omega=-1}$ diverges.

\end{appendix}


\end{document}